\documentclass[twocolumn,amsmath,amssymb,floatfix,showpacs]{revtex4}

\usepackage{graphicx, color, dcolumn, bm}
\usepackage[normalem]{ulem}

\begin{document}

\title{Inverse cascade behavior in freely decaying two-dimensional fluid 
       turbulence}
\author{P.D. Mininni$^{1,2}$ and A. Pouquet$^2$}
\affiliation{$^1$ Departamento de F\'\i sica, Facultad de Ciencias Exactas y
         Naturales, Universidad de Buenos Aires and IFIBA, CONICET, Ciudad 
         Universitaria, 1428 Buenos Aires, Argentina. \\
             $^2$ NCAR, P.O. Box 3000, Boulder, Colorado 80307-3000, U.S.A.}
\date{\today}

\begin{abstract}
We present results from an ensemble of 50 runs of two-dimensional 
hydrodynamic turbulence with spatial resolution of $2048^2$ grid points, 
and from an ensemble of 10 runs with $4096^2$ grid points. All runs in 
each ensemble have random initial conditions with same initial integral scale, 
energy, enstrophy, and Reynolds number. When both ensemble- and time-averaged, 
inverse energy cascade behavior is observed, even in the absence of external 
mechanical forcing: the energy spectrum at scales larger than the 
characteristic scale of the flow  follows a $k^{-5/3}$ law, with negative 
flux, together with a $k^{-3}$ law at smaller scales, and a positive flux of 
enstrophy. The source of energy for this behavior comes from the modal 
energy around the energy containing scale at $t=0$. The results shed some 
light into connections between decaying and forced turbulence, and into 
recent controversies in experimental studies of two-dimensional and 
magnetohydrodynamic turbulent flows.
\end{abstract}
\pacs{47.27.-i,47.32.C-,47.65.-d}
\maketitle

\section{Introduction}

Forced and freely decaying flows are different, as far as long-time 
energetics is concerned. While forced flows can develop both ``direct'' 
and ``inverse cascades'' \cite{Kraichnan80,Frisch} (transfer of ideally 
conserved quantities to smaller or larger scales with constant flux), 
freely decaying flows can at most display ``selective decay'' 
\cite{Matthaeus80,Ting86} (the faster decay of one invariant relative 
to another, when the system has two or more ideally conserved quantities 
and one of them condenses at large scales in the ideal case). These two 
processes are in practice related: selective decay results from the 
transfer of one of the invariants to short wavelengths where the 
dissipation coefficients are more effective.

For flows with only direct transfer of ideal invariants, e.g., 
three-dimensional (3D) hydrodynamic flows, the distinction between freely 
decaying and forced flows is not that important. It is well known that 
if time-averaged around the time of maximum dissipation, the spectrum 
of the ideal invariant develops in the decaying case the same scaling as 
in forced cases. It is also known that the shape of the spectrum during 
the decay is preserved under certain conditions \cite{Kolmogorov41}, 
resulting in a self-similar decay. The concept of ``direct cascade'' is 
then often applied to these flows indistinctly of the nature of the 
mechanisms sustaining the turbulence.

However, for flows with inverse transfer of an invariant, the distinction 
between freely decaying and forced flows remains, and dates back to before 
the introduction of concepts such as selective decay. In two-dimensional 
hydrodynamic flows, on the one hand, Kraichnan considered the forced case 
\cite{Kraichnan67} and predicted an inverse cascade of energy with a 
direct cascade of enstrophy. On the other hand, Batchelor \cite{Batchelor69} 
considered the direct transfer of enstrophy in decaying flows, obtaining 
the long discussed $\sim k^{-1}$ enstrophy spectrum (see, e.g., 
\cite{Bracco00,Dmitruk05,Dritschel07,Lindborg10}). The reasons for the 
distinction in this case are clear: as explained by Kraichnan, the 
possibility of an inverse cascade depends on the relative strengths of 
the non-linear interactions between scales. It is natural then to assume 
that if the input of energy is not sustained for sufficiently long times, 
the inverse cascade cannot develop and that in the flow only the direct 
cascade, with an increase of the energy containing scale, will be observed 
(note that the increase of this scale is however distinguishable from the 
three-dimensional hydrodynamic case).

The main aim of this work is to study under what conditions, in a 
decaying flow, behavior associated with an inverse cascade (approximately 
constant negative flux, and the corresponding self-similar spectrum with the 
appropriate scaling law) arises. The motivation to identify such behavior 
is twofold: On the one hand, phenomena observed in solar wind turbulence 
and space plasmas are often loosely associated with an inverse cascade of 
magnetic helicity 
\cite{Christensson01,Demoulin09a,Demoulin09b}. In many of these cases 
(as, e.g., the solar wind) the large scale dynamics can be well 
approximated by a decaying magnetohydrodynamic (MHD) flow. Although 
forced MHD flows develop inverse cascades (of magnetic helicity in three 
dimensions \cite{Pouquet76}, and of the square vector potential in two 
\cite{Pouquet78}), based on the previous arguments results obtained from 
simulations or theory of forced turbulence do not strictly apply to the 
decaying case. On the other hand, recent experiments of decaying soap 
film flows (see \cite{Kellay02} for a review) showed in some cases positive 
third-order velocity structure function \cite{Belmonte99}, inverse 
energy transfer for scales larger than the injection scale \cite{Rivera03}, 
thickness behaving as a passive scalar with Kolmogorov-like scaling 
\cite{Greffier02}, and in one case even a disputed $\sim k^{-5/3}$ energy 
scaling \cite{Gharib89}, results that are considered indications of 
inverse cascades in forced two-dimensional flows, but are puzzling in 
the decaying case.

\section{Numerical procedure}

The systems discussed above are complex, with 
small-scale plasma effects in the case of space flows \cite{Mininni03} and 
with deviations from two-dimensionality \cite{Couder89} and finite-size 
effects \cite{Belmonte99,Clercx00} in the case of soap films. These effects 
are important to explain the observations (see, e.g., the discussions in 
\cite{Belmonte99,Greffier02}). With a more humble objective in mind, we 
take these problems as motivation, and focus our attention on differences 
that may arise in simulations and experiments because of the differences 
in the procedures followed to obtain turbulent data. To this end, we 
restrict ourselves to a much simpler system: that of two-dimensional 
hydrodynamic flows in the absence of external forces, freely decaying 
in periodic boxes.

In high resolution numerical simulations, because of the computational 
cost, seldom an ensemble of simulations with different initial conditions 
is done. Instead, only one simulation is done, and ergodicity is assumed 
in order to obtain statistical quantities by averaging in space. 
Experiments of decaying turbulence are characterized instead by single- 
or multiple-point measurements of quantities downstream for very long 
times. The multiple-point measurements can be at similar stages of the 
evolution, or at different stages (e.g., when measuring at different 
distances downstream from the place where turbulence is generated). The 
procedure actually results on an average over multiple realizations of 
decaying flows.

\begin{figure}
\includegraphics[width=8.5cm]{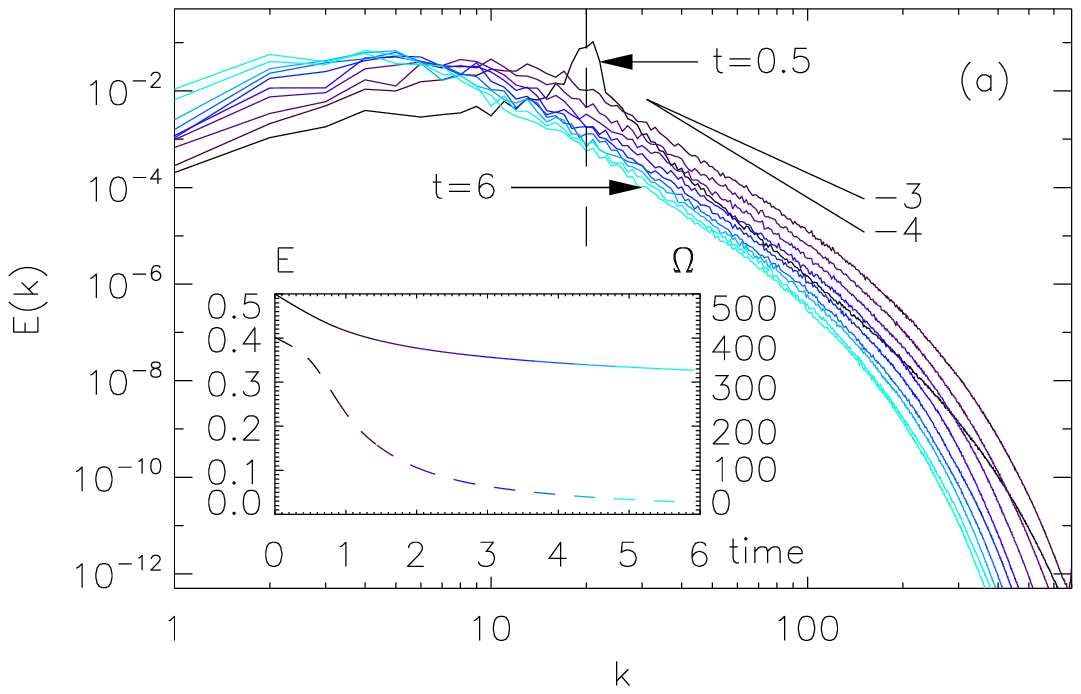}
\includegraphics[width=8.5cm]{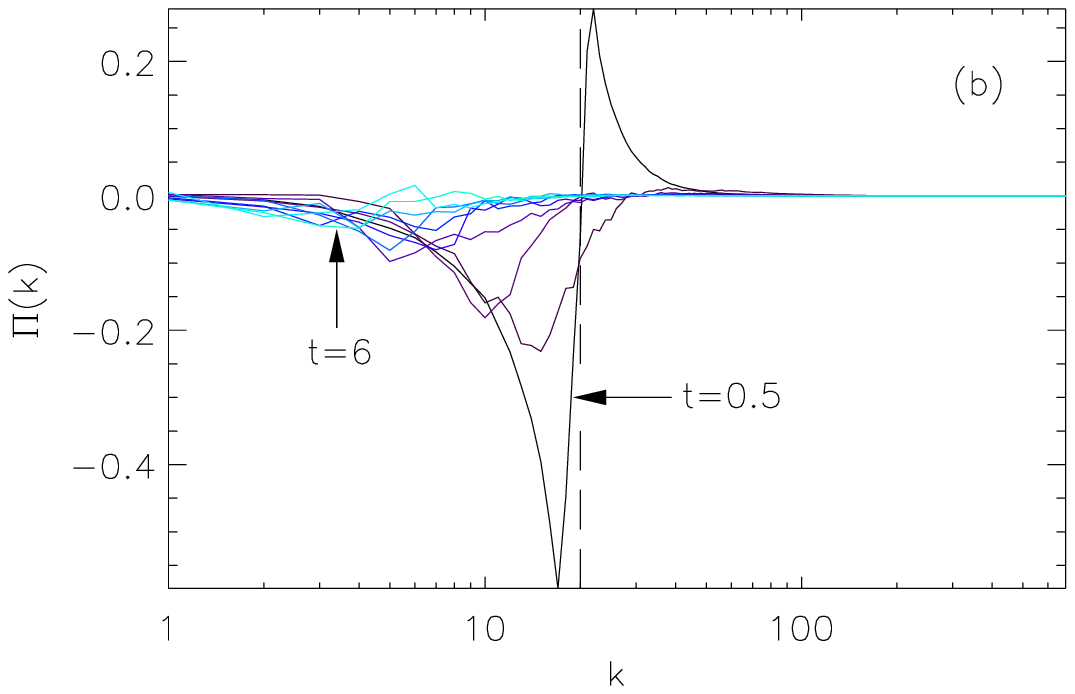}
\includegraphics[width=8.5cm]{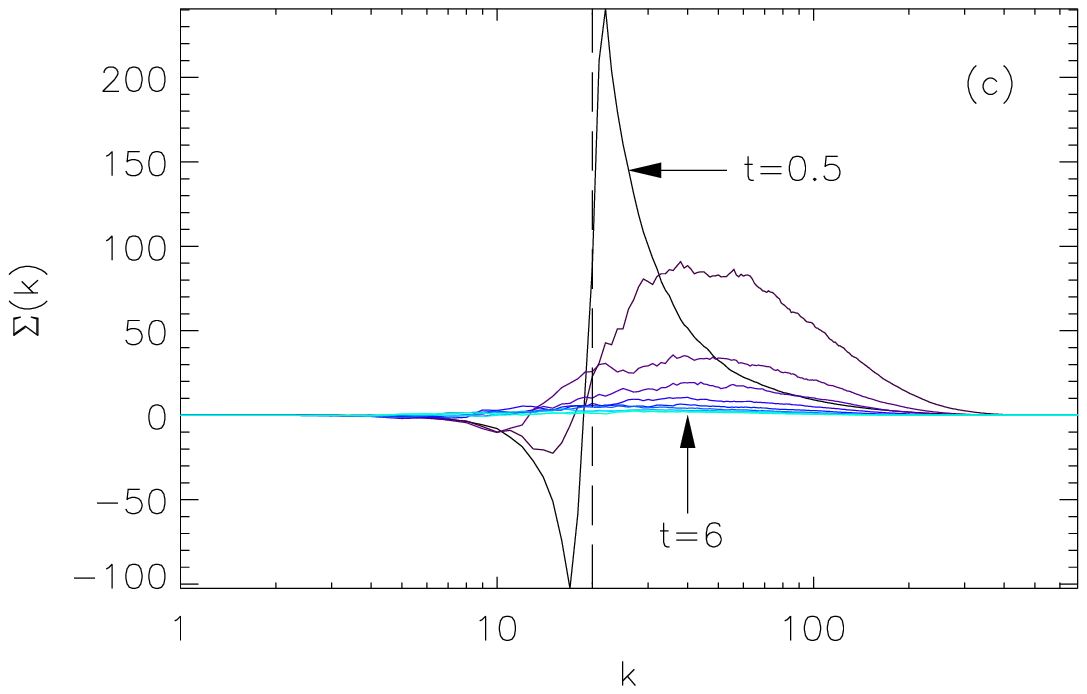}
\caption{(Color online) Time evolution of the energy spectrum (a), 
    energy flux (b), and enstrophy flux (c) in a single $2048^3$ simulation, 
    from $t=0.5$ (black) to $t=6$ (light gray or light blue). Slopes in 
    the energy spectrum are indicated as references. The curves 
    corresponding to $t=0.5$ and to $t=6$ are indicated in all panels by 
    arrows, and the vertical dashed lines indicate the initial energy 
    containing wave number $k_0$. In (b) and (c), note the displacement 
    to smaller wave numbers of the minimum of energy flux, and to larger 
    wave numbers of the maximum of enstrophy flux. The inset in (a) shows 
    the time evolution of the energy (solid line) and of the enstrophy 
    (dashed line) in this run, 
    with the color changing with time following the colors used for the 
    different curves in the spectrum and fluxes.}
\label{fig:instene} \end{figure}

To mimic such a procedure in our simulations, we performed 50 
two-dimensional simulations of the Navier-Stokes equations on a regular 
grid of $2048^2$ points. Initial conditions were a random superposition 
of harmonic modes between wave numbers $k=18$ and $22$, with the spectrum 
peaking at $k_0=20$. Viscosity was $\nu = 2.5 \times 10^{-4}$ in all these 
runs, and the box had length $2\pi$. The initial r.m.s.~velocity $U_{r.m.s.}$ 
in all runs was $1$, corresponding to a turn-over time of 
$\tau_{NL} = L_0/U_{r.m.s.} = 2\pi/20 \approx 0.3$ ($L_0=2\pi/k_0$). As a 
result, the runs only differed by their random initial phases.

To study the effect of scale separation between the initial 
energy-containing scale and the box size, a second set of 10 simulations 
was done with a resolution of $4096^2$ grid points. The initial random 
excitation with $U_{r.m.s.}=1$ was placed between $k=30$ and $34$, with the 
spectrum peaking at $k_0=32$. The initial integral scale of the flow was 
thus decreased by a factor $1.6$, and the initial turnover time was 
$\tau_{NL} \approx 0.2$. The viscosity in these runs was 
$\nu = 9.9 \times 10^{-5}$, and as in the previous dataset, the runs in 
these dataset only differed in the choice of initial phases.

The equations were integrated up to $t=6$ using a parallelized 
pseudo-spectral code, with the $2/3$-rule for dealiasing, and 
second-order Runge-Kutta to evolve in time \cite{Gomez05}. Neither 
friction nor forcing were employed. Note that although the spatial 
resolutions considered in this study are moderate, the computational cost 
of computing the whole dataset (the 50 $2048^2$ runs and the 10 $4096^3$ 
runs) is equivalent to that of computing a $16384^2$ simulation of 2D 
turbulence \cite{Boffetta12}. 

\section{Results}

\begin{figure}
\includegraphics[width=8.5cm]{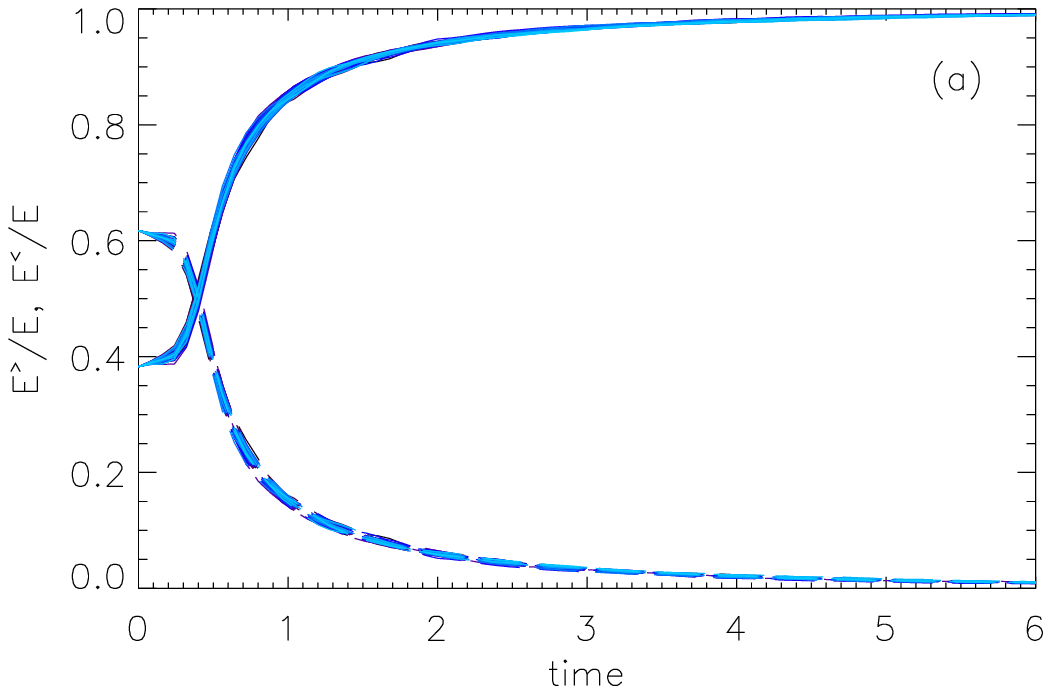}
\includegraphics[width=8.5cm]{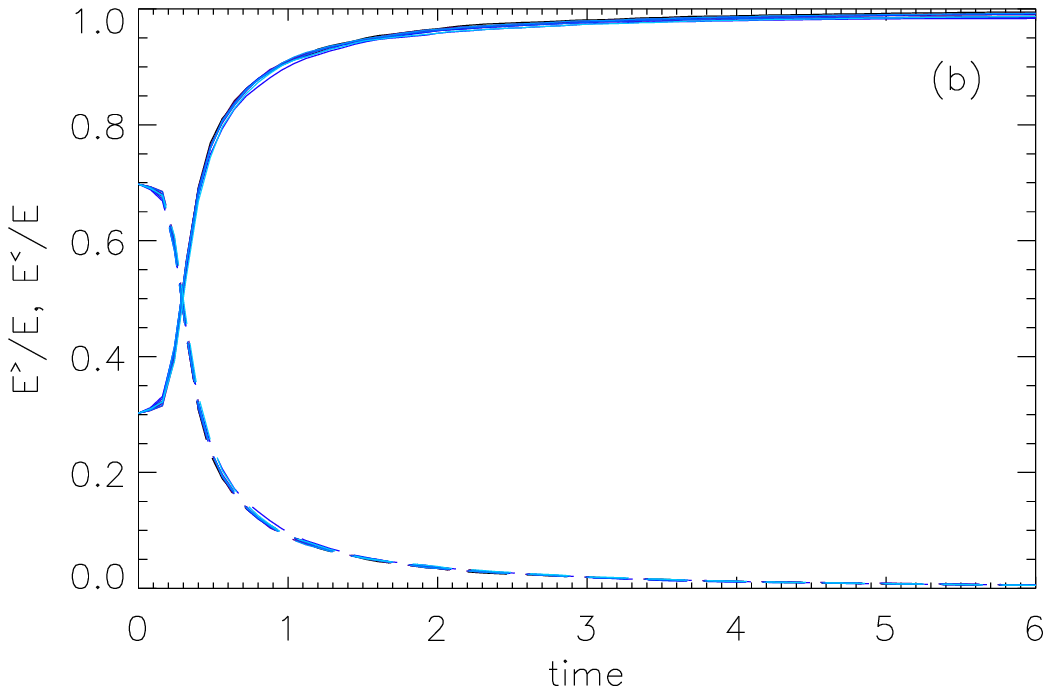}
\caption{(Color online) Time evolution of $E^</E$ (solid) and $E^>/E$ 
    (dashed) (resp., ratio of the energy at large and at small scales 
    to the total energy) as a function of time for all runs in the 
    set of simulations with $2048^2$ grid points (a), and with 
    $4096^2$ grid points (b).}
\label{fig:Elarge} \end{figure}

Since the first set has more realizations (50), resulting in more reliable 
ensemble averages, we focus first on this set and compare at the end with 
the second dataset to identify the effect of increasing resolution. 
Figure \ref{fig:instene} shows the time evolution of the energy spectrum, 
of the energy flux, and of the enstrophy flux in one of the $2048^3$ 
simulations, i.e., for a single realization. At large wave numbers the 
energy spectrum develops a direct enstrophy cascade range with an energy 
spectrum steeper than $\sim k^{-3}$ but shallower than $\sim k^{-4}$. Such 
spectra have been reported before, and for details we refer the reader to 
the discussion in \cite{Lindborg10} and references therein. We simply 
note that, given the separation of scales in this run 
$k_{max}/k_0\approx 34$, with $k_{max}=N/3$ the maximum wavenumber, one 
does not expect to be able to resolve sufficiently the small-scale range. 
Recent simulations of forced two-dimensional turbulence, with more scale 
separation between $k_0$ and $k_{max}$, indicate a clear $\sim k^{-3}$ energy 
scaling at small scales \cite{Vallgren11}, as expected from Batchelor 
phenomenology (see also \cite{Kraichnan80,Clercx00}). Since this is not the 
main purpose of this paper, we just want to stress that the resulting 
inertial range is compatible with moderate Reynolds numbers runs done in 
the past.

We are interested here instead in the spectrum at wave numbers smaller 
than the initial energy containing wave number $k_0$. At those wave 
numbers, the peak of energy moves towards smaller wave numbers as time 
evolves, leaving to its left an energy spectrum again consistent with the 
direct enstrophy range. The fluxes are in agreement with this picture: 
the energy flux has a negative peak that moves towards smaller wave numbers 
with time, but has no identifiable range with approximate constancy. On the 
other hand, the enstrophy flux develops a wide quasi-constant range 
starting at $t\approx 1$, and as its amplitude decreases with time, this 
range can still be identified.

\begin{figure}
\includegraphics[width=8.5cm]{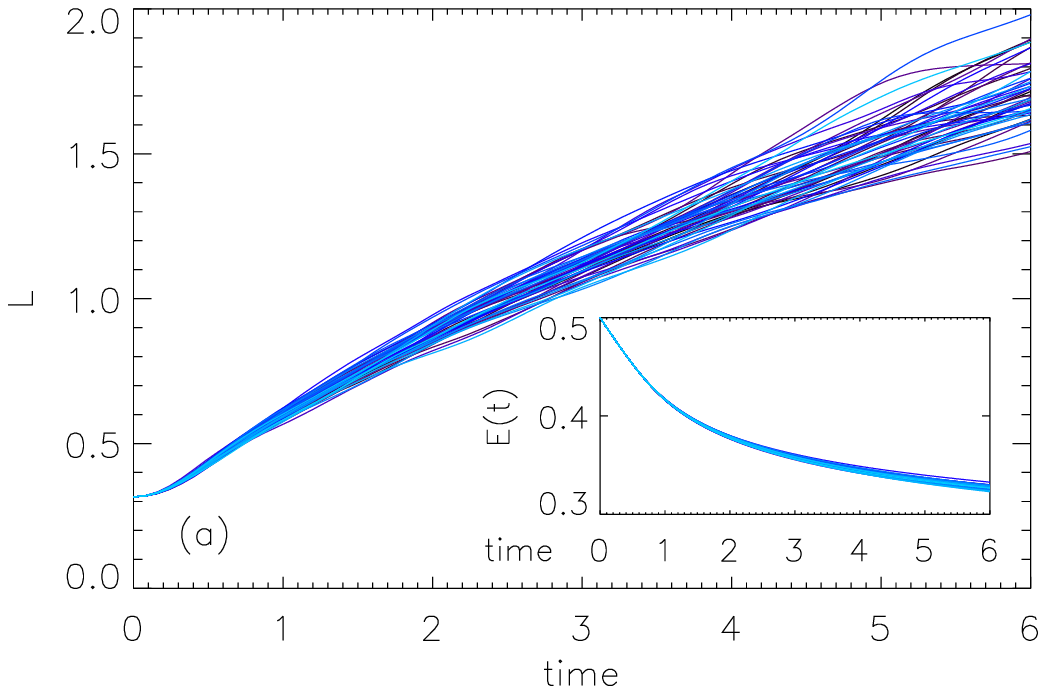}
\includegraphics[width=8.5cm]{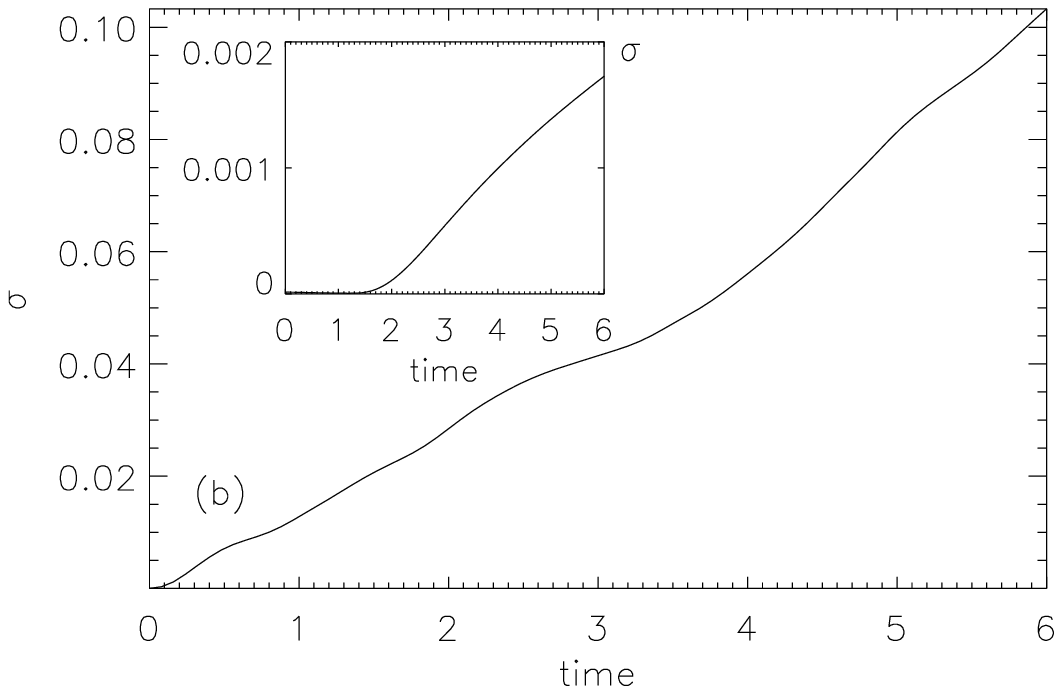}
\caption{(Color online) (a) Time evolution of the integral scale for all the 
    $2048^2$ runs; note the dispersion around the mean evolution. The inset 
    shows the time evolution of the energy for all runs, which are very 
    close. (b) Time evolution of the standard deviation $\sigma$ in the 
    integral scale of all the $2048^2$ runs. The inset shows the standard 
    deviation in the energy in all the $2048^2$ runs as a function of time.}
\label{fig:integral} \end{figure}

The displacement of the peak of the energy spectrum towards smaller 
wave numbers can be associated with the well-known increase of the flow 
integral length $L$ as the energy decays. However, it does not 
automatically follow that the increase of $L$ is the sign of an 
inverse cascade. Indeed, the integral length scale also grows in a 
three-dimensional flow as is well known, due to the preservation of 
large scale correlations in the von K\'arm\'an-Howarth equation 
\cite{Davidson,Ishida06} with spectra $\sim k^{+n}$, with $n$ either 
equal to two or four, and also due to the eddy-noise created by the 
beating of two interacting small scales. As the peak of the energy 
spectrum in three-dimensional turbulence moves to larger scales, the 
amplitude of the peak decreases, being modulated by the $\sim k^{+n}$ 
spectra. On the other hand, the growth of the integral scale in an 
inverse cascade is accompanied by another phenomenon, that of the 
growth of energy at large scales.

Unlike freely decaying three-dimensional turbulence, the energy at 
large scales in the two-dimensional runs increases substantially as 
time evolves. This can be seen in Fig.~\ref{fig:Elarge}, that shows 
the energy at large scales (wavenumbers smaller than $k_0$)
\begin{equation}
E^< = \sum_{k=1}^{k_0-1} E(k) ,
\end{equation}
and the energy at small scales (wavenumbers larger than $k_0$)
\begin{equation}
E^> = \sum_{k=k_0}^{k_{max}} E(k) ,
\end{equation}
normalized by the total energy as a function of time. In both runs, 
the small-scale energy decreases rapidly, while the large-scale energy 
grows. When not normalized by the total energy, $E^<$ still grows and 
then slowly decays as a result of viscous forces (with a slower decay 
in the $4096^2$ run). In the absence of transfer of energy towards 
large scales (i.e., without negative energy flux), $E^<$ would only 
decrease in time.

The time evolution of the integral scale for all $2048^2$ runs is shown 
in Fig.~\ref{fig:integral}. In spite of the fact that the runs only 
differ by their initial phases, there is a large dispersion in the 
time evolution of $L$, more so than for the energy (see the inset of 
Fig.~\ref{fig:integral}(a)). The actual dispersion in the evolution of 
$L$ and $E$ in the different runs is also quantified in 
Fig.~\ref{fig:integral}(b), by means of the standard deviation $\sigma$. 
For both quantities, $\sigma$ first grows exponentially (at very early 
times), and later seems to follow an approximate linear growth with time. 
From pioneering works on predictability of two-dimensional turbulence 
\cite{Lorenz69,Leith71,Leith72} (see also \cite{Boffetta97,Boffetta01} 
for recent studies), we can expect differences in the initial conditions 
to grow first at the initial energy containing scale (resulting in the 
early exponential phase), and later to propagate towards larger scales 
if an inverse cascade develops (see \cite{Boffetta01} for a numerical 
study). In that regime, the time it takes for the differences to 
propagate is that of the turnover time, which if a Kolmogorov spectrum 
is assumed, results in linear growth of the error with time 
\cite{Lorenz69,Boffetta01}. Note that the integral scale $L$ is 
obtained from the energy spectrum weighting the most energetic 
wavenumbers, and as such more deviation can be expected than in 
the case of the total energy.

\begin{figure}
\includegraphics[width=8.5cm]{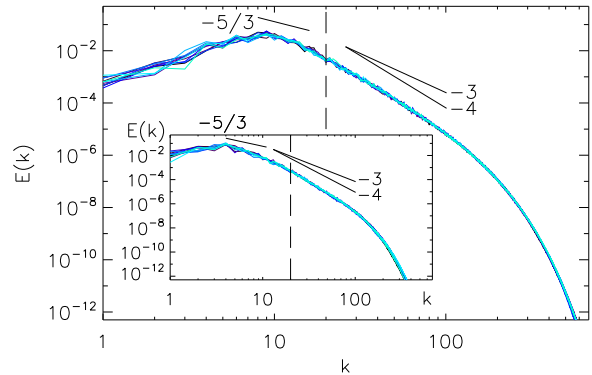}
\caption{(Color online) Energy spectra for 10 $2048^2$ runs at $t=1.5$
    (no averaging performed here). Slopes 
    are indicated as references. The vertical dashed line corresponds to 
    the initial energy containing wave number $k_0$. The inset shows the 
    same ten spectra at $t=6$.
    Note that the large-scale spectra look like $\sim k$, which could 
    be interpreted as eddy-noise in two-dimensions (see \cite{Fox09}). 
    However, unlike three-dimensional turbulence, the amplitude of the 
    peak of the energy spectrum is larger than its initial value 
    (at $t=0$) at the same wavenumber.}
\label{fig:spectrum} \end{figure}

\begin{figure}
\includegraphics[width=8.5cm]{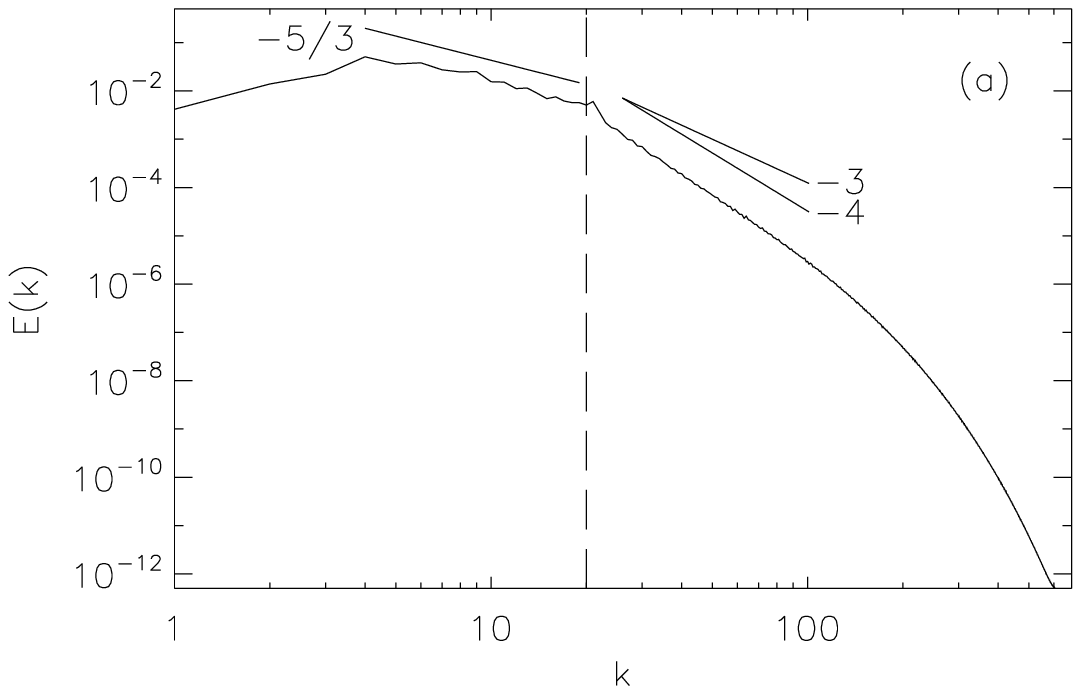}
\includegraphics[width=8.5cm]{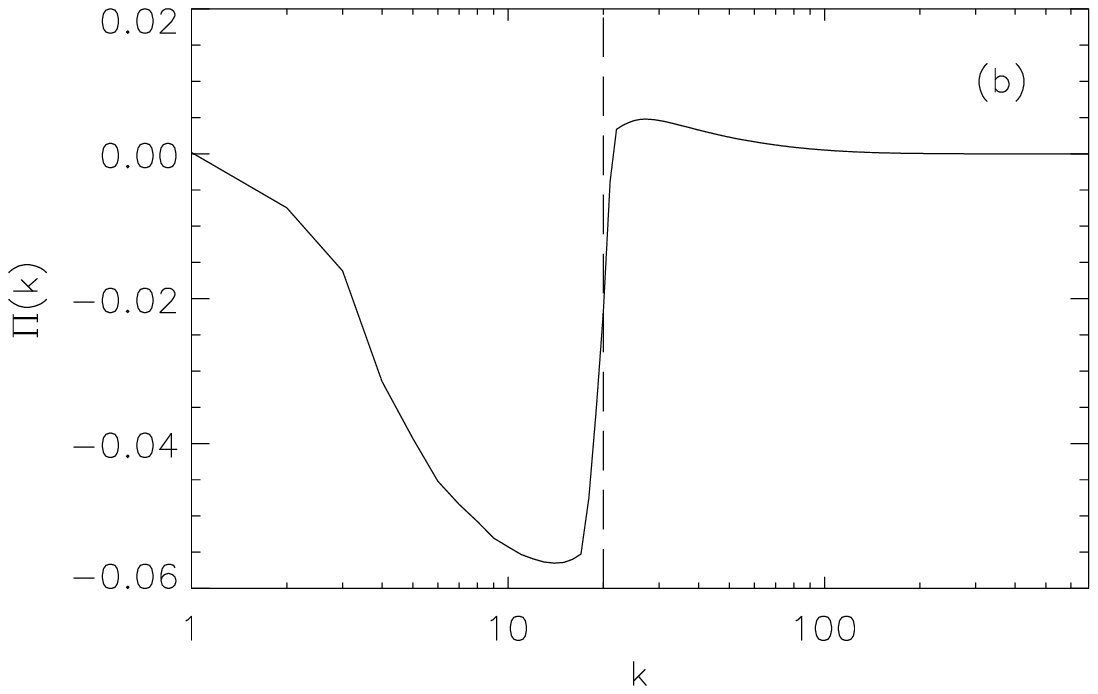}
\includegraphics[width=8.5cm]{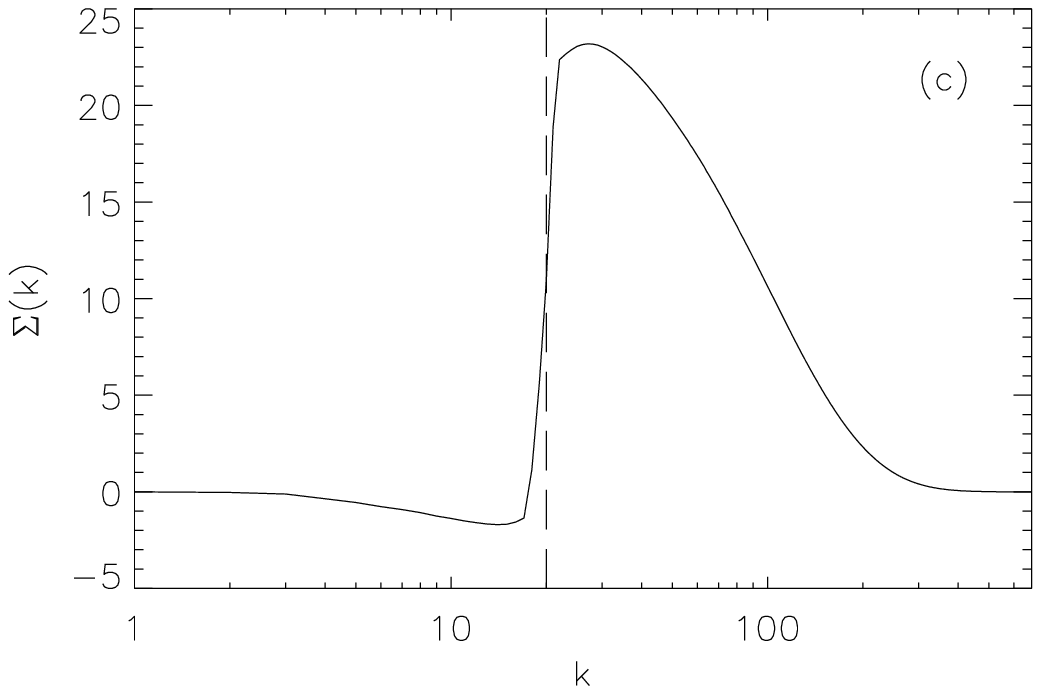}
\caption{Time- and ensemble-averaged energy spectrum (a), energy flux (b), 
    and enstrophy flux (c) over the 50 $2048^2$ simulations and from $t=0.5$ 
    to $t=6$. Slopes in (a) are indicated as references. The vertical dashed 
    lines corresponds to the initial energy containing wave number $k_0$.}
\label{fig:ensemble} \end{figure}

Although the standard deviation observed in Fig.~\ref{fig:integral} can be 
expected in an ensemble, numerical simulations often deal with only one of 
the realizations. What are the implications of the dispersion in the 
integral length (and energy) in the different simulations at a given 
time? Figure \ref{fig:spectrum} shows the energy spectrum at $t=1.5$ 
and at $t=6$ for 10 randomly picked $2048^2$ runs. At $t=1.5$, a narrow 
range seems to emerge at large scales, with a slope shallower than 
$\sim k^{-3}$, between the energy containing scale and the enstrophy 
range. Exploration of the energy flux (not shown) indicates that its 
ensemble average may display a short range with approximate
constancy. Moreover, at later time (see, e.g., the inset in 
Fig.~\ref{fig:spectrum}), this shallower range in the energy spectrum 
increases in width and  moves to smaller wave numbers.

Based on this result, we show in Fig.~\ref{fig:ensemble} the time- and 
ensemble-averaged energy spectrum, energy flux, and enstrophy flux for 
the set of simulations with $2048^2$ grid points. The average is 
computed over the 50 simulations and from $t=0.5$ to $t=6$. Now, a 
wide $\sim k^{-5/3}$ spectrum clearly emerges at wave numbers smaller than 
the initial energy-containing wave number, followed by the enstrophy range 
at larger wave numbers. The energy flux now also displays a wider range 
in which it remains negative and non-negligible (note that negative energy 
flux corresponds to positive third-order longitudinal velocity structure 
function). Moreover, a clear distinction can be made between the range 
with negative energy flux and negligible enstrophy flux, and the range 
with positive enstrophy flux and negligible energy flux. Finally, note 
that, dimensionally, $\Sigma\sim k^2_0 |\Pi|$, which is well verified by 
the data; in other words, the observed energy and enstrophy fluxes 
(respectively to large and small scales) are comparable, when properly 
normalized using the characteristic length scale of the flow at $t=0$.

\begin{figure}
\includegraphics[width=8.5cm]{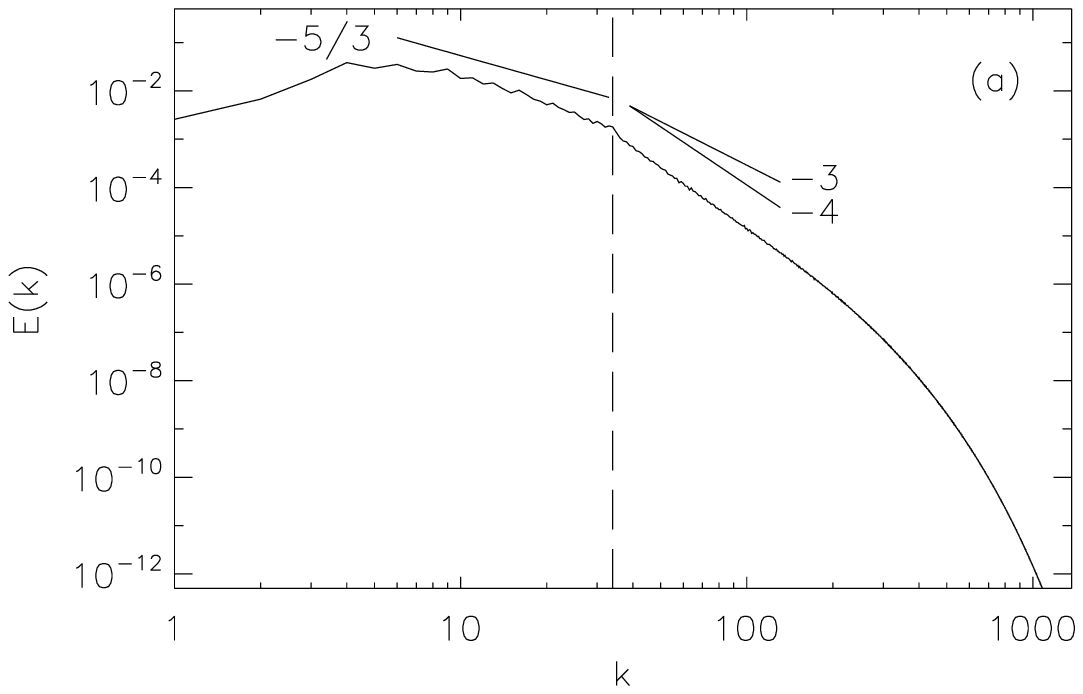}
\includegraphics[width=8.5cm]{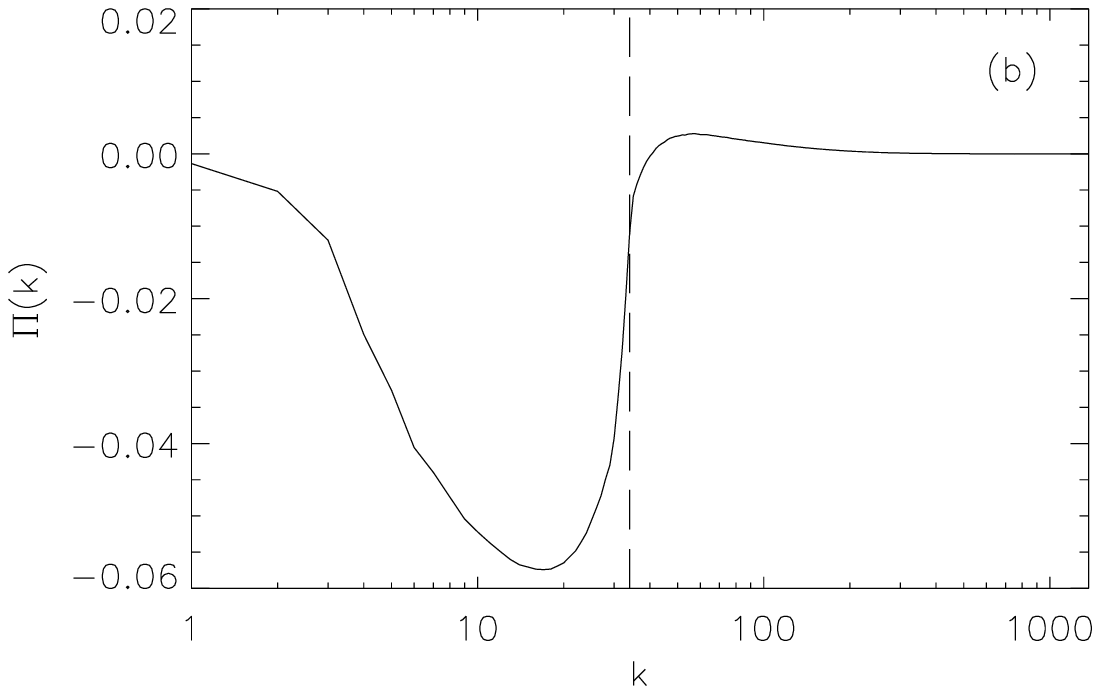}
\caption{Time- and ensemble-averaged energy spectrum (a) and energy flux (b) 
    over the 10 $4096^2$ simulations and from $t=0.5$ to $t=6$. Slopes in 
    (a) are indicated as references. The vertical dashed lines corresponds 
    to the initial energy containing wave number $k_0$.}
\label{fig:ensemble4096} \end{figure}

The same time- and ensemble-averages are shown in Fig.~\ref{fig:ensemble4096} 
for the 10 runs with $4096^2$ grid points (as the scale separation between 
initial perturbation and dissipation scale is roughly the same in this set 
of runs as in the runs with $2048^2$ points, the flux of enstrophy is 
similar as the one in Fig.~\ref{fig:ensemble} and is not shown). The 
larger scale separation at large scales allows for the development of a 
slightly wider  $\sim k^{-5/3}$ spectrum.

The results in Figs.~\ref{fig:ensemble} and \ref{fig:ensemble4096} 
indicate that after time- and ensemble-averaging, well-known features 
of the inverse energy cascade in forced two-dimensional hydrodynamic 
turbulence can be identified in decaying flows. The $\sim k^{-5/3}$ 
energy spectrum emerges as an envelope to the evolution of the 
large-scale energy spectrum, which instantaneously and for each run 
only shows a clear direct enstrophy range. It is interesting to point 
out that time- and ensemble-averages of freely decaying two-dimensional 
turbulence were used before (see \cite{Dritschel08}). There, a 
spectrum for the enstrophy $\sim k$ (i.e., $E(k) \sim k^{-1}$) was 
observed at scales larger than the initial energy containing scale. Although 
this indicates a growth of energy at large scales, the spectrum is 
shallower than the one reported here, probably because in that study 
much later times were considered when studying the decay (over 100 
turnover times).

\section{Discussion}

Experiments, as well as observations of space flows, do not allow for such 
a tight control of initial r.m.s.~velocities and initial injection scales 
as in the present simulations, which kept these quantities the same in 
all the initial conditions. It can be conceived that as these initial 
quantities are allowed to vary within a certain percentage, the dispersion 
of the integral scale and of the energy at a later time during the decay 
will become larger. This, in a much larger ensemble of runs, may allow 
for a $\sim k^{-5/3}$ energy spectrum to arise without time averaging, 
as each realization at a given time may correspond to different stages 
of the decay.

Although in the applications discussed in the introduction other 
effects may be playing a role to explain the observations, it is however 
remarkable that a link between the behavior of forced and decaying flows 
was missed for such a long time. We offer a possible explanation for this. 
In a time when computers allow us to study larger and larger Reynolds 
numbers, and many studies are focused on achieving the largest possible 
Reynolds number, the present study points at the need to use computer 
power also to perform longer runs, and ensembles of runs (as routinely 
done in climate studies) to be able to compare on equal footing with 
experiments and theory, and to validate some of the usual assumptions in 
the field, such as ergodicity (in the sense that time- and 
spatial-averages can be seamlessly interchanged when comparing experiments 
and simulations), and, as was our focus here, distinctions between 
forced and decaying flows.

It is worth recalling in this context that the statistical mechanics 
of ideal (unforced) flows, which have a signature of condensate of 
energy at large scales, were the basis on which Kraichnan's 
conjecture of the inverse cascade was made, so that it is not 
necessarily surprising to find such a behavior in the decay run as well, 
since it stems from the nature of two-dimensional nonlinear triadic 
interactions, and the constraints of conservation. Indeed, the triads 
responsible for the inverse cascade of energy are present in freely 
decaying flows, and whether they can act or not is a problem of 
energetics (i.e., whether their amplitude is large enough, and can be 
sustained for sufficiently long times). It is therefore important to 
point out that it has been shown that a $\sim k^{-5/3}$ energy 
spectrum is the minimally steep spectrum that is able to support an 
inverse cascade of energy \cite{Tran07} (i.e., a shallower spectrum cannot 
sustain the inverse cascade). The behavior observed in these freely 
decaying simulations can thus only be observed for a certain period of 
time, as the energy is spread over more wave numbers and the inverse 
energy flux decreases with time (see Fig.~\ref{fig:instene}). In the 
forced case, the external forcing supplies the required energy to sustain 
the flux in time and to give a persistent $\sim k^{-5/3}$ energy 
spectrum.

Interestingly, recently other problems were found in fluid dynamics 
in which particular triads allowed for transient or for sustained 
inverse cascades. Such is the case, e.g., of three-dimensional 
helical flows restricted to triadic interactions with only the same 
sign of helicity in all interacting modes \cite{Biferale11}. In this 
later case, the inverse cascade was sustained in time instead of 
being transient, as it resulted from an artificial amplification in the 
simulation of particular triads.

Other similarities between decaying flows and the forced Kraichnan's 
inverse energy cascade may arise, and are left for future work. In 
particular, intermittency of the large scales, the effect on the 
structures, and the equivalent behavior in MHD can be studied and may 
be of interest for many applications.

{\it Computer time was provided by the National Center for Atmospheric 
Research, which is sponsored by the National Science Foundation. The 
authors acknowledge support from NSF-CMG grant No.~1025183, and PDM 
acknowledges support from PICT grants No. 2011-1529 and 2011-1626.} 

\bibliography{ms}

\end{document}